\documentclass[prl,twocolumn,floatfix,epsfigs,nofootinbib]{revtex4}
\usepackage{amssymb,amsmath,amsfonts}
\usepackage{graphics,dcolumn,bm,fleqn,epic,eepic,float}
\usepackage{graphicx} 
\usepackage{epstopdf} 
\usepackage{bm} 
\usepackage{subfigure}
\usepackage{color} 

\begin{document}
\title{Why are most COVID-19 infection curves linear? 
}

\author{Stefan Thurner$^{1,2,3}$, Peter Klimek$^{1,2}$, Rudolf Hanel$^{1,2}$, }

\affiliation{%
  $^1$~Section for Science of Complex Systems, Medical University of
  Vienna, Spitalgasse 23, 1090 Vienna, Austria}
   \affiliation{%
  $^2$~Complexity Science Hub Vienna, Josefst\"adterstrasse 39, 1080 Vienna, Austria}
 \affiliation{%
  $^3$~Santa Fe Institute, Santa Fe, NM 87501, USA}
\date{\today}

\begin{abstract} 
Many countries have passed their first COVID-19 epidemic peak. Traditional epidemiological models describe this as a result of non-pharmaceutical interventions that pushed the growth rate below the recovery rate. In this new phase of the pandemic many countries show an almost linear growth of confirmed cases for extended time-periods. This new containment regime is hard to explain by traditional models where infection numbers either grow explosively until herd immunity is reached, or the epidemic is completely suppressed (zero new cases). Here we offer an explanation of this puzzling observation based on the structure of contact networks. We show that for any given transmission rate there exists a critical number of social contacts, $D_c$, below which linear growth and low infection prevalence must occur. Above $D_c$ traditional epidemiological dynamics takes place, as e.g. in SIR-type models. When calibrating our corresponding model to empirical estimates of the transmission rate and the number of days being contagious, we find $D_c\sim 7.2$. Assuming realistic contact networks with a degree of about 5, and assuming that lockdown measures would reduce that to household-size (about 2.5), we reproduce actual infection curves with a remarkable precision, without fitting or fine-tuning of parameters. In particular we compare the US and Austria, as examples for one country that initially did not impose measures and one that responded with a severe lockdown early on. Our findings question the applicability of standard compartmental models to describe the COVID-19 containment phase. The probability to observe linear growth in these is practically zero.
\end{abstract}

\date{May 22, 2020}

\maketitle

\section{Introduction}
Textbook knowledge of epidemiology has it that an epidemic event comes to a halt 
when herd immunity in a population is reached \cite{textbook,herd}. 
Herd immunity levels depend on the disease. For influenza it is within the range of 33-44\% of the population \cite{herd1}
for Ebola it is 33-60\% \cite{herd2}, for measles 92-95\% \cite{herd2}, and for SARS levels between 50-80\% 
are reported \cite{herd3}. 
For the current COVID-19 outbreak it is expected to be in the range of 29-74\% \cite{herd4,herd5}. 
On way towards herd immunity, textbook knowledge teaches, the number of infected increases 
faster-than-linear (in early phases even exponentially) as long as the effective reproduction number is larger than $1$.
Once this threshold is passed, the daily increments in the numbers of infected starts to decrease until it drops to zero \cite{textbook,sir}.
Combining these two growth phases gives the characteristic S-shaped infection curves.

The COVID-19 outbreak shows a very different picture, however. 
Several countries have  clearly passed the maximum of the epidemic, 
and are converging towards zero new cases per day. 
None of these countries are even close to herd immunity. 
In Austria at the hight of the pandemic, a  population-wide representative PCR study  
showed that only about 0.3\% of the population were tested positive \cite{sora}. 
Similarly, in Iceland in a random-population screening 
the prevalence of positively tested was found to be 0.8\% \cite{iceland}.
Clearly, the COVID-19 outbreak is far from the uncontrolled case as many countries have 
implemented non-pharmacautical interventions to reduce infection rates \cite{amelie}.

\begin{figure}[tb]
	\centering
	 \includegraphics[width=0.51\columnwidth]{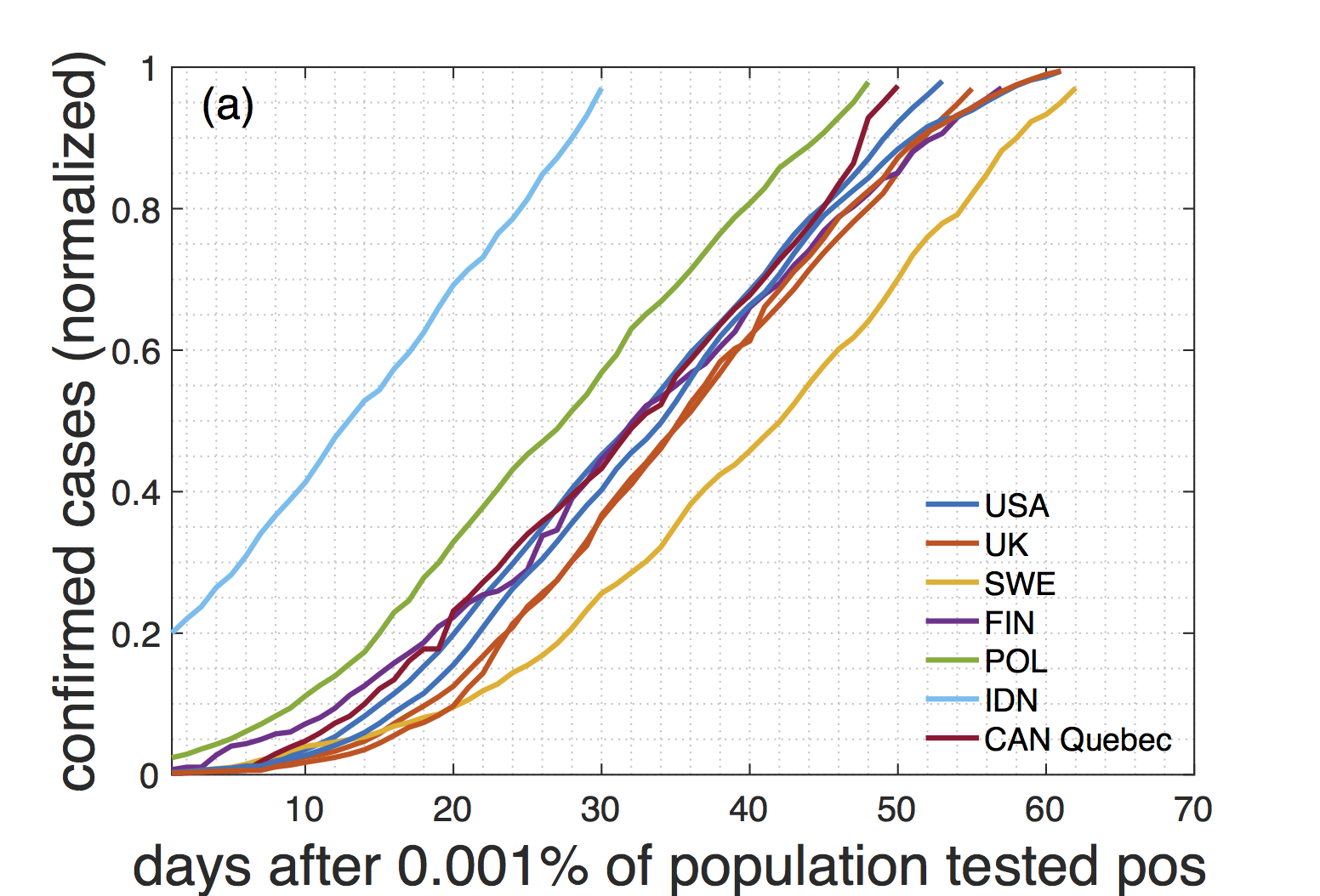}
	\hspace{-0.5cm} \includegraphics[width=0.51\columnwidth]{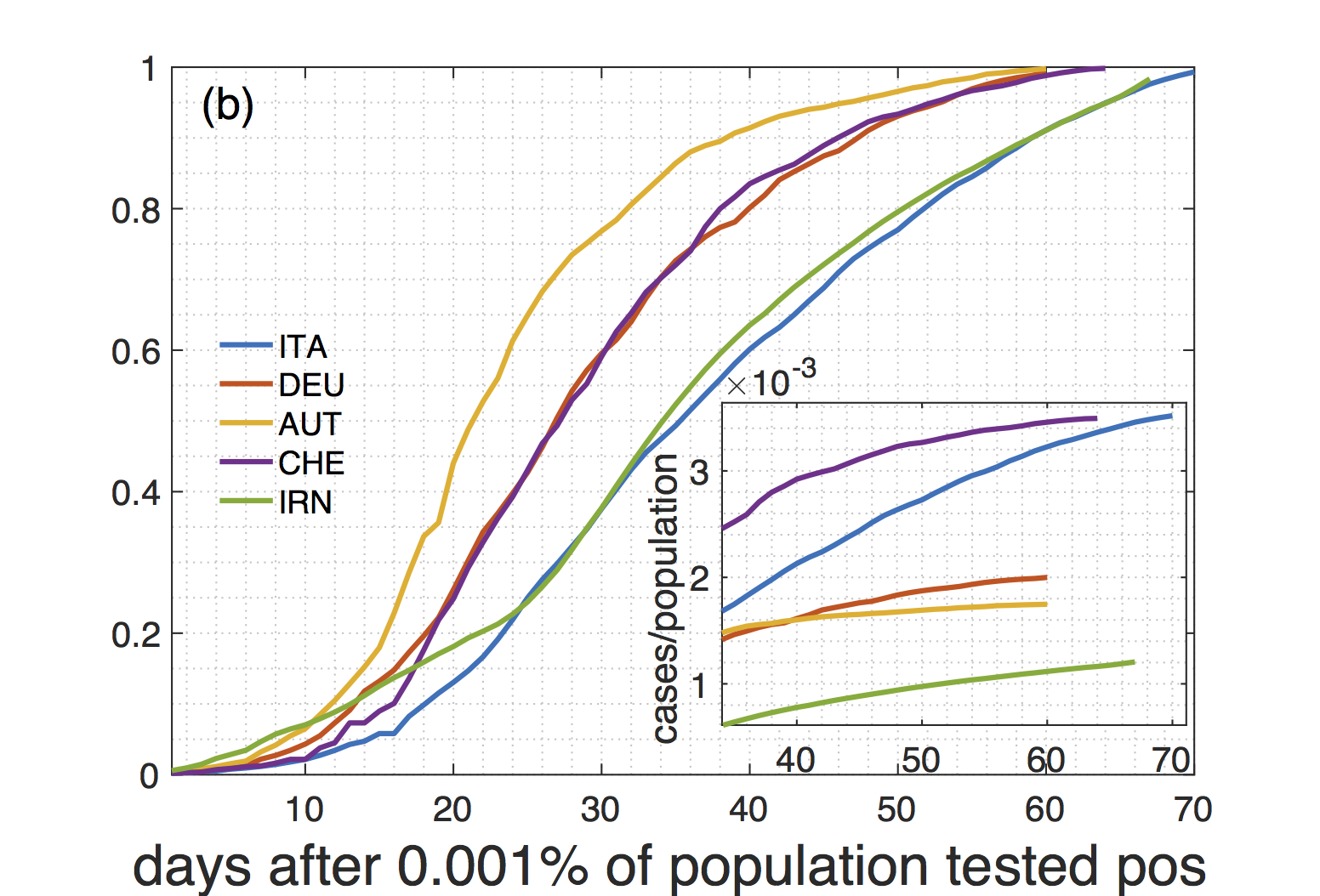}
	\caption{Cumulative numbers of positively tested cases normalized to the last day (May 8, 2020). 
	Countries, even though many followed radically different  strategies in response to the pandemic, 
	seem to belong to one of three groups: 
	(a) Countries with a remarkably extended linear increase of the cumulated number of positively tested cases, 
	including the US, UK and Sweden. 
	(b) Countries with an extended  linear increase that tend to level off and enter a regime with a smaller slope. 
	Inset shows extended regime after the peak (cases per population size). 
	}
	\label{fig:linear}
\end{figure}

Maybe the most striking observation in the COVID-19 infection curves is that they 
exhibit linear growth for an extended time interval quite in contrast to the S-shaped curves expected from 
epidemiological models. 
For wide range of countries regardless of size, demographic and ethnic composition, or geo-location, 
this linear growth pattern is apparent even by a plain-eye inspection of the number of positive cases, 
see e.g. \cite{JHU}. 
In Fig. \ref{fig:linear} (a) we show infection curves (number of confirmed positive cases)  
for the US, UK, Sweden, Finland, Poland, Indonesia, and a province of Canada. 
Clearly, after a short initial exponential phase, infection curves are practically linear for several weeks. 
For many other examples, see \cite{JHU}.
Many countries that implemented non-pharmaceutical interventions in response to the COVID-19 crisis \cite{amelie}, 
show a different pattern. They also show an extended linear growth, however, infection curves 
tend to bend and level off in response to the implemented measures, see Fig. \ref{fig:linear} (b) . 
The extent of the linear regime depends on the onset of the measures \cite{amelie}. 
Many countries that are in the early phase of the pandemic are in still (May 8, 2020) show the initial almost exponential growth, see SI Fig. \ref{SIfig:linear}. 
According to basic epidemiological concepts, growth patterns with extended linear regions 
are not to be expected. They can only be observed if the infection growth rate equals the recovery rate, 
giving an effective reproduction number, $R(t)$, of $1$.
Chances of observing such a behaviour over an extended period of time in a country are extremely tiny, 
let alone in several countries. 
Mathematically speaking, linear growth is basically a measure zero solution in compartmental models.

The basic question of this paper is to clarify the mechanism that keeps $R(t) \sim1$. 
In classical SIR-type \cite{sir} models there are no terms that explicitly peg $R(t)$ to 1; see SI. 
A simple explanation could be a limiting capacity of availability of test kits. 
If the number of tests is limited on every day only and assuming a fixed ratio of 
confirmed cases per test, linear growth in the number of positively tested would be the consequence. 
However, most European counties, even though experiencing initial difficulty with testing capacity, 
have, by now, enough tests. 

The rationale underlying social distancing efforts is that they lead to a reduction of contacts which essentially 
makes the social network sparser \cite{amelie}. Infections occur if 
(i) there is a social interaction between an infected and susceptible person, and 
(ii) this contact is intense enough to lead to a disease transmission.
For instance, given a basic reproduction number of $R_0 \sim 3$, we effectively reach herd immunity if 
two out of three contacts are avoided.
Still, this does not yet explain linear growth as a slight increase or decrease in contact probabilities 
would again lead to a faster-than-linear growth or suppression, respectively.
Network density alone can not explain persistent linear growth.

In classic epidemiology network effects have long been ignored in favour of analytical tractability \cite{may}.
In that case epidemiological models can be formulated as differential equations, assuming that 
every person in principle can infect any other. This is called the well mixed, or mean-field approximation; See SI. 
However, that fact that networks matter in epidemiology is been recognized for almost 2 decades and lead 
to extremely relevant contributions, such as the dependence of vaccination thresholds on network topology, see e.g. 
\cite{vespignani}. 
Classic contributions such as \cite{newman, moreno} were able to incorporate network topology into analytically 
solvable SIR models. There it is possible to solve the SIR model in terms of outbreak size and epidemic size, 
however, no focus was put on the details of infection curves below the epidemic limit. 
When dealing with structured networks it might well be that the mean-field approximation does no-longer hold, 
and details of the networks start to become crucial. 

Since social networks are key to understand details of epidemic outbreaks, 
what do they look like? The answer is highly non-trivial since social networks are hard to define.
In terms of network topology, it became clear that they are neither pure random graphs, small-world 
networks, nor are they purely scale-free. They are of a more involved structure, 
including multi-level organization \cite{szell}, weak-ties between communities \cite{kerteszPNAS} and 
temporal aspects that suggest a degree of fluidity, however, with stable social cores \cite{socialNWpnas}. 

Here we try to understand the origin of the extended linear regime in infection curves, 
as currently observed in the number of positively tested cases in COVID-19 pandemic across many countries. 
To this end we solve the SIR model on a simple social network and report a hitherto 
unobserved transition from linear growth to S-shaped infection curves.
We show that for a given transmission rate there exists a critical degree below 
which linear growth is expected and above which the model reproduces the classical SIR results. 
Below this critical degree the mean-field approximation starts to fail. 
For the underlying social networks we use a Poissonian small-world network that 
tries to capture several empirical facts, including 
a heterogenous number of social links (degree),    
the small-world aspect, 
the fact that people tend to live in small groups (families), 
that these groups overlap, 
and that work and leisure relations can link distant groups;  see Methods.  
The framework allows us to model a lockdown as a change of social networks 
with a high degree to one with a degree that characterizes the members of a household.  
Based on data on household size in the EU \cite{eurostat},  
empirical estimates on how long individuals are contagious, 
and on transmission (or attack) rates we are able to calibrate the model to real countries.  
In particular we compare the situation in the USA and Austria. 
These countries differ remarkably in size and the measures taken in response to the COVID-19 pandemic \cite{amelie}.
While Austria imposed a lockdown relatively early on in combination with a number of other measures, 
the US has implemented measures hesitantly with the consequence that the situation was  ``not under control'', as
Dr A. Fauci, an advisor to the Trump administration, has put it on May 12, 2020 \cite{fauci}.
The model reproduces the real infection curves to a remarkable degree. 
All parameters are empirically motivated, there are no fitted parameters involved. 

\section{Model}

{\bf Model dynamics.}
We assume that there are $N$ individuals connected by social links. 
If $i$ and $j$ are connected, $A_{i,j}=1$,  if they are not $A_{i,j}=0$. 
As a toy model for social networks we use a so-called small-world network 
with average degree $D$ and shortcut probability $\epsilon$; see Methods.
The small-world aspect allows us to model transmission between local groups and 
``superspreaders'' \cite{Kupferschmidt}.
As in a SIR model, every individual is in one of three possible states, 
susceptible (S), infected (I), and recovered (R). 
If an individual is infected it will infect its susceptible neighbours with a per-day transmission probability, $r$.
This means that on every single day the probability of passing the infection to a susceptible neighbour is $r$,  
which is sometimes called the microscopic spreading rate \cite{moreno}. 
Once a person is infected it stays infectious for $d$ consecutive days. After this the person can no longer infect
others, and is called recovered. Once recovered the state will no longer change. 
The  update rules of the corresponding model are:
\begin{itemize}
\item initialize all nodes as susceptible. Select $N_{ini}$ nodes randomly and change their state to infected, 
\item at every timestep $t$, find all infected and infect their susceptible neighbors  with probability $r$,
\item set all infected nodes that have been infected for more than $d$ timesteps to recovered,
\item proceed to the next timestep until the dynamics comes to a halt. 
All nodes are now either recovered or susceptible.
\end{itemize}
At every timestep (day),  $t$, we count the number of new cases, $C(t)$;
the infection curve of positive cases, $P(t)$, is the cumulative sum of $C(t)$.
The model parameters are related to those of the SIR model (see SI) by 
$\gamma=1/d$, and $\beta=rD/N$. If the underlying network 
fulfils the conditions necessary for the mean-field approximation, 
$C(t)$ corresponds to $R(t)$ up to a timeshift of $d$. 

\begin{figure}[tb]
	\centering
	  \includegraphics[width=0.96\columnwidth]{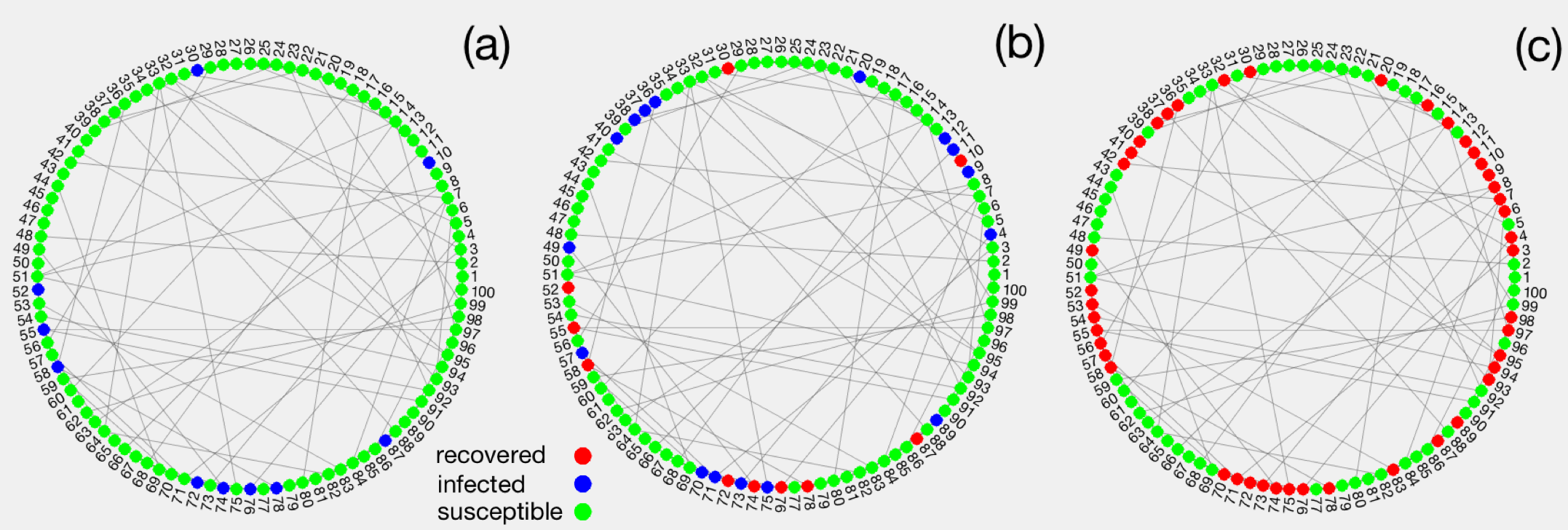} 
	   \includegraphics[width=1.0\columnwidth]{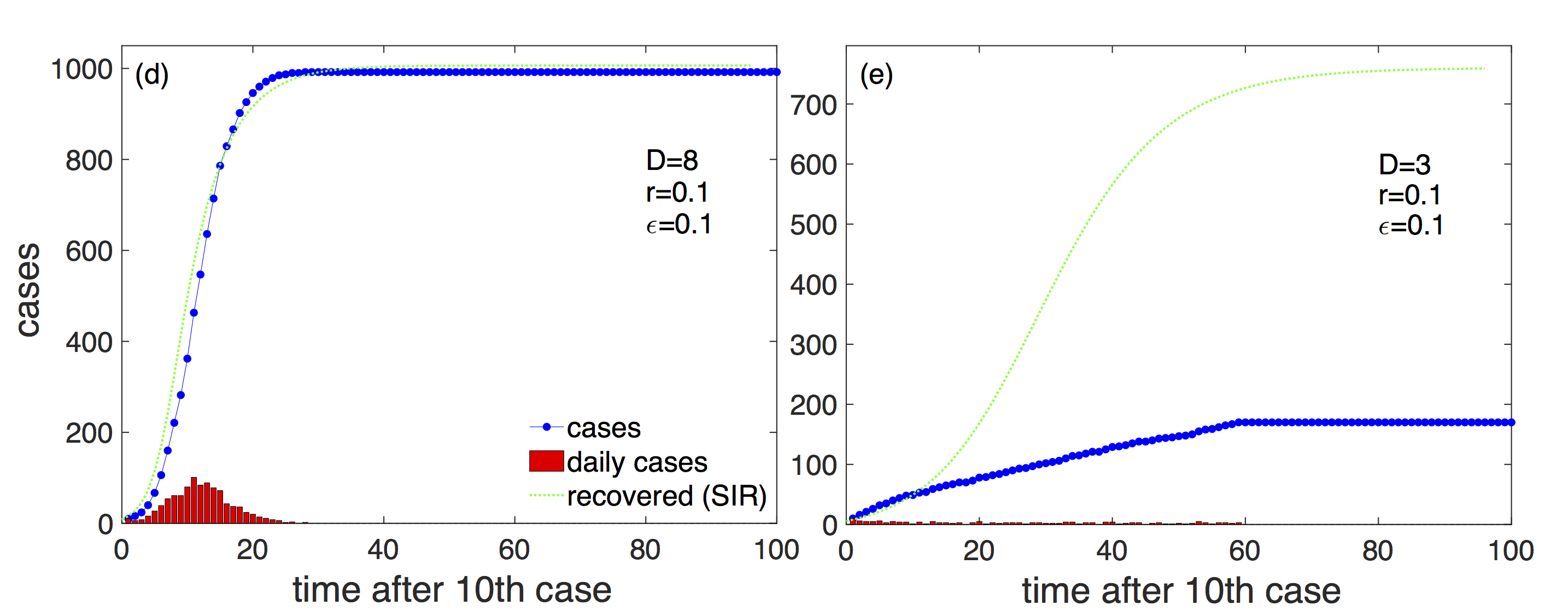}
		\caption{Schematic demonstration of the model. Nodes are connected in a Poissonian small-world network. 
		Locally close neigbors resemble the family contacts, long links to different regions represent contacts 
		to others, such as people at work. 
		(a) Initially, a subset of nodes are infected (yellow), most are susceptible (green). 
		(b) At every timestep, infected nodes spread the disease to any of their neighbors with probability $p$. 
		After $d$ days infected nodes turn into `recovered' and no longer spread the disease.   
		(c) The dynamics ends when no more nodes can be infected and all are recovered.  
    (d) Infection curve $P(t)$ (blue dots) for the model on a dense Poissonian small-world network, $D=8$.  
     The daily cases (red) first increase and then decrease. For comparison, we show the recovered cases, $R(t)$, 
     of the corresponding SIR model with $\gamma=1/d$, and $\beta=rD/N$ (green). The mean-field conditions 
     are obviously justified to a large extent.
    (e) Situation for the same parameters except for a lower average degree, $D=3$.
    The infection curve now increases almost linearly; daily increases are nearly constant for a long time. The dynamics 
    reaches a halt at  about 17\% infected. The discrepancy to the SIR model (green) is now obvious. 
	}
	\label{fig:nw}
\end{figure}

\section{Results}
{\bf Infection dynamics.} 
We demonstrate the model schematically in Fig. \ref{fig:nw} (a)-(c). 
In the limit of large degree $D$ and large $\epsilon$ 
the model should approximately fulfill the mean-field conditions
and should be  close to a classical SIR model. 
This is seen in Fig. \ref{fig:nw} (d) where the trajectory of an infection curve, $P(t)$, is 
shown (blue dots) for a network of 1000 nodes with a degree of $D=8$, $\epsilon=0.1$, 
a period of contagiousness of $d=6$ days, 
and an transmission rate of $r=0.1$; 10 nodes were infected at the start.  
The situation closely resembles the solution of the recovered, $R(t)$, the SIR model with 
$\gamma=1/d$, and $\beta=rD/N$, shown as the dotted green line.
Note that a timeshift of $-d$ days is necessary to compare $P(t)$ and $R(t)$. 
The daily cases (red) increase, reach a peak and decrease. 
The typical exponential initial phase in $P(t)$ is seen, immediately followed by 
a quick relaxation of growth until the plateau forms at the herd immunity level (in this example at 98\%).  

The infection curve, $P(t)$, changes radically when the degree of the network is lowered to $D=3$ 
(all other parameters kept the same); see Fig. \ref{fig:nw} (e). 
Clearly, it increases almost linearly for a remarkable timespan, which is marked in contrast to the SIR expectation (green line). 
The situation already resembles the situation of many countries. 
Once the system converged to its final state, only about 17\% of nodes were infected, 
which is far from the expected  (SIR) herd immunity level of about 77\%. 

The change of the infection curve from the  S-shaped to a linear behavior 
is clearly a network effect and indicates that the mean-field assumptions might be violated. 
To understand this better we next study the parameter dependence more systematically. 

\begin{figure}[tb]
	\centering
	 \includegraphics[width=0.5\columnwidth]{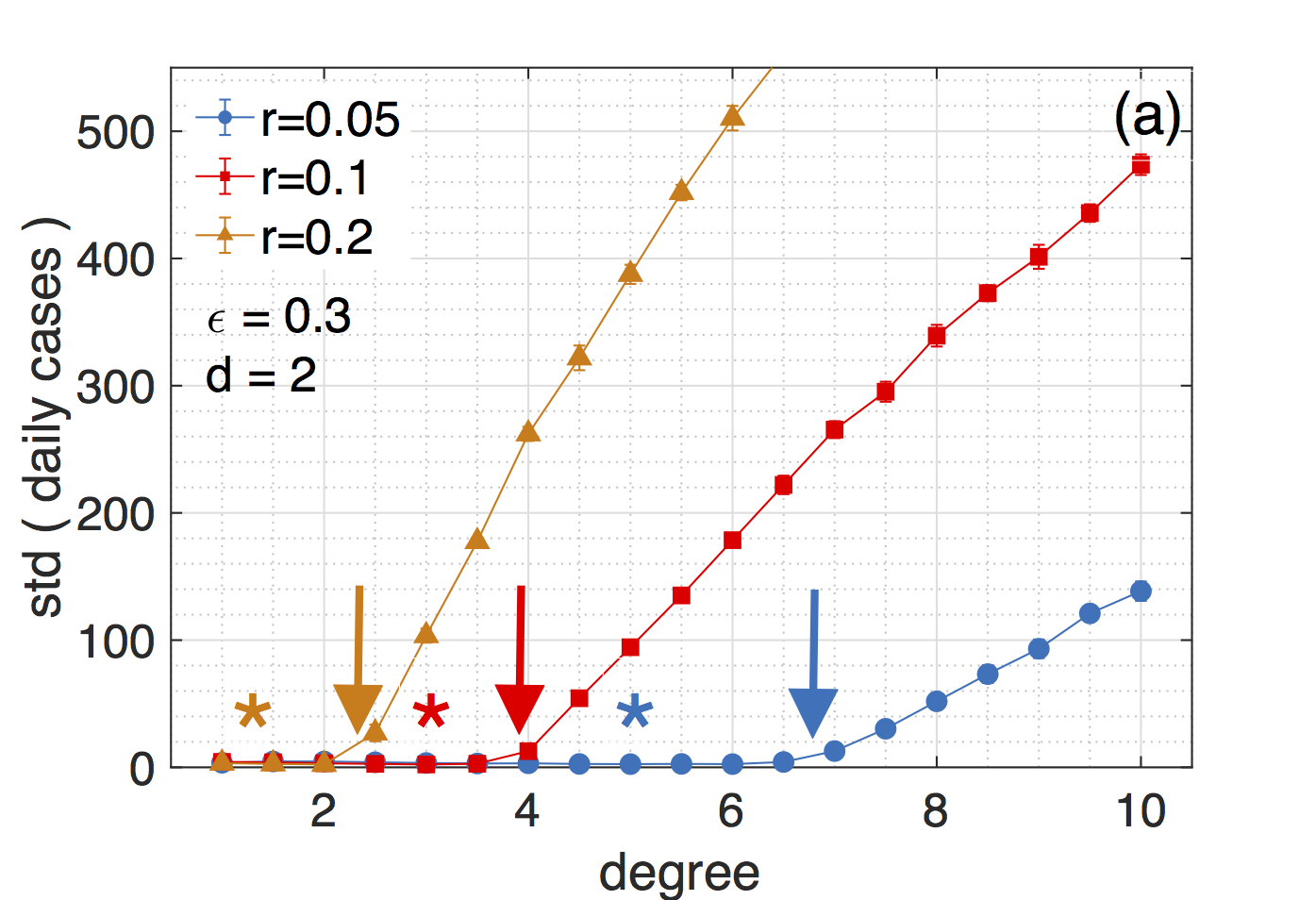} 
	\hspace{-0.5cm} \includegraphics[width=0.5\columnwidth]{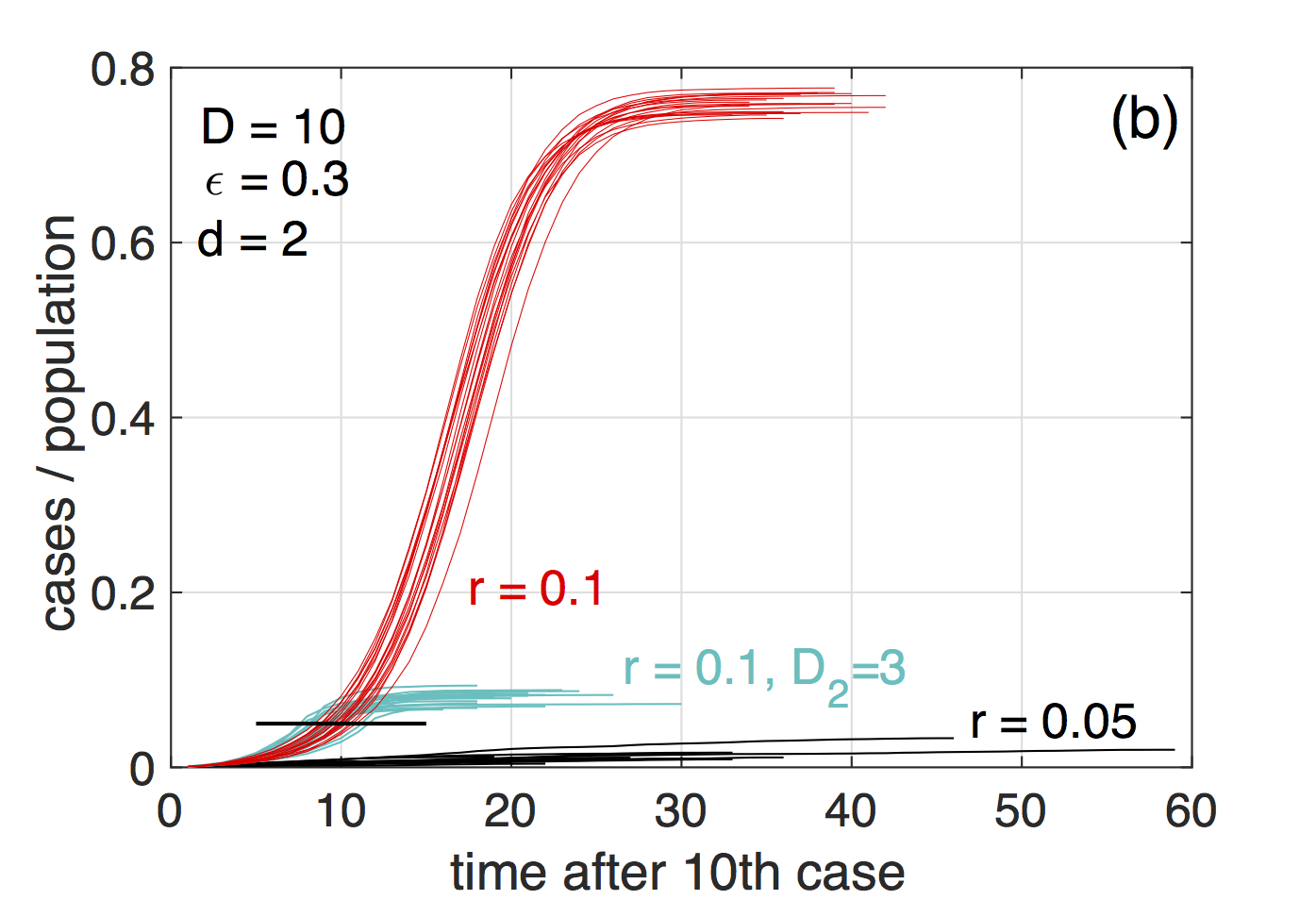} 
		\caption{
		(a) Order parameter for the transition from linear to S-shaped infection curves as a function of degree, 
		$D$, for transmission rates $r=0.05$ (blue), $0.1$ (red), and $0.2$ (orange). The transition happens at the 
		critical points, $D_c$, where the order parameter starts to diverge from zero (arrows); see Table \ref{Table1}.
		(b) Infection curves (20 realizations) for three scenarios for a network with $D=10$. 
		Scenario red: at an transmission rate of $r=0.1$ we see S-shaped curves reaching herd immunity at about 75\%.
		Scenario black: for the same network with a lower transmission rate, $r=0.05$, we fall below the critical degree $D_c $and 
		consequently observe linear growth; note the convergence of infected at levels of $1-4$\%, 
		which are very much below herd immunity (75\%). 
		Scenario blue (lockdown): we start with the same network with $r=0.1$, 
		as in the red scenario. After 5\% of the population (black bar)
		is infected there is a lockdown, that changes the network to one of degree $D_2=3$, from one day to the next.
		The S-shaped growth immediately stops and levels off at about 10\% infected. 
		Other parameters: $d=2$, $\epsilon=0.3$, and $N=10,000$.  10 initially infected.	
	}
	\label{fig:order}
\end{figure}

{\bf Parameter dependence \& phase transition.} 
We are interested to see if there is a critical degree, $D_c$, below which the infection curve 
is (quasi) linear, whereas for $D>D_c$ it assumes the S-shape.
For this we define an appropriate ``order parameter'', ${\cal O}$, able to distinguish 
linear from S-shape growth, namely the standard deviation of daily increments of infected people; see Methods.
In Fig. \ref{fig:order} (a) we show this order parameter as a function of the degree, $D$, of the network 
for three transmission rates $r=0.05$, $0,1$, and $0.2$ 
(obtained as averages over 10 independent realizations with randomly selected 10 initially infected). 
It is clear that at specific (critical) degrees, $D_c$, the order parameter switches from (close to) zero to larger values. 
The position of the critical  degrees depend on the parameter settings (arrows). 
It decreases with the transmission rate $r$; 
while for $r=0.05$ we find $D_c=6.6$, for $r=0.1$ it is $D_c=3.9$, and for $r=0.2$, we have $D_c=2.3$. 
The critical degree also decreases with the parameters $\epsilon$ and $r$.
For more parameter settings, see Table \ref{Table1}, and SI Fig. \ref{SIfig:X} in the SI. 
The asterix denote the degree, $D_{\rm sir}$, at which the SIR model would show a linear curve, $D_{\rm sir}=1/dr$.
 
\begin{table}[ht]
\caption{Critical degree, $D_c$, for a range of parameter values for 
the transmission rate, $r$, rewiring probability, $\epsilon$, and duration of infectiousness, $d$. 
$N=10,000$.}
\begin{center}
\begin{tabular}{c c c c c c }

			 &               & r=0.015 &  $r=0.05$ &  $r=0.1$  & $r=0.2$  \\
\hline
$\epsilon=0.1$   	&  $d=2$   &		&    12.5	&	7.4		&	4.3	\\
			&  $d=4$   &		& 	7.9	&	4.8		&	3.0	\\
		   	&  $d=6$   &		& 	5.7	&	3.5		&	2.4	\\
\hline
$\epsilon=0.3$   	&  $d=2$   &		&    12.2	&	6.3		&	3.3	\\
			&  $d=4$   &		& 	6.6	&	3.9		&	2.3	\\
		   	&  $d=6$   &		& 	4.7	&	2.9		&	1.9	\\
		   	&  $d=14$   &	7.2	& 	2.7	&	1.9		&	1.5	\\
\hline
$\epsilon=0.5$   	&  $d=2$   &		&    11.7	&	5.9		&	3.2	\\
			&  $d=4$   &		& 	6.0	&	3.3		&	2.2	\\
		   	&  $d=6$   &		& 	4.4	&	2.7		&	1.8	\\
\end{tabular}
\end{center}
\label{Table1}
\end{table}

We checked that the position of the critical degrees is relatively robust under the size of the network, 
and the topology. We find that for $N=1000$ and 10 initially infected, the critical degrees are practically at the same locations.
Regarding the topology, we implemented a standard small-world network with a fixed degree. Also here, results 
are practically identical; see SI in SI Fig. \ref{SIfig:SW}. 
 
The existence of critical degrees signals the presence of a 
hitherto overlooked `transition' between linear and 
S-shape growth that is most likely due to the fact that the well-mixed or mean-field assumption breaks 
down below $D_c$. 
To illustrate the dependence of this transition on the transmission rate, 
Fig. \ref{fig:order} (b) shows 20 realizations of 
model infection curves for a network with $D=10$ at a rate of $r=0.1$ (red). The curves were obtained for 
20 different initial conditions in the choice of the 10 initially infected nodes.
One observes typical S-shape curves reaching herd immunity at about 75\%.
Note that $R(t\to\infty)$ of the SIR model reaches about 80\%.
For the same network with a lower transmission rate of $r=0.05$, which is well 
below the critical degree we are in the linear growth domain (blue). 
The maximum of infected reaches levels of only $1-4$\%, which are drastically lower than SIR 
herd immunity with $R(t\to\infty)\sim$  15\%.
The 20 green infection curves depict a ``lockdown'' scenario: 
we start with the same network with $r=0.1$ (red). On the day when 5\% of the population is infected (black bar) 
a lockdown is imposed which means that effectively the social network changes from one day to the next. 
We model this by switching to a Poissonian small-world network with a lower degree, $D_2=3$. All other parameters 
are kept equal. S-shape growth immediately stops and final infection levels of about 10\% are obtained. 

\begin{figure}[tb]
	\centering
	 \includegraphics[width=1.1\columnwidth]{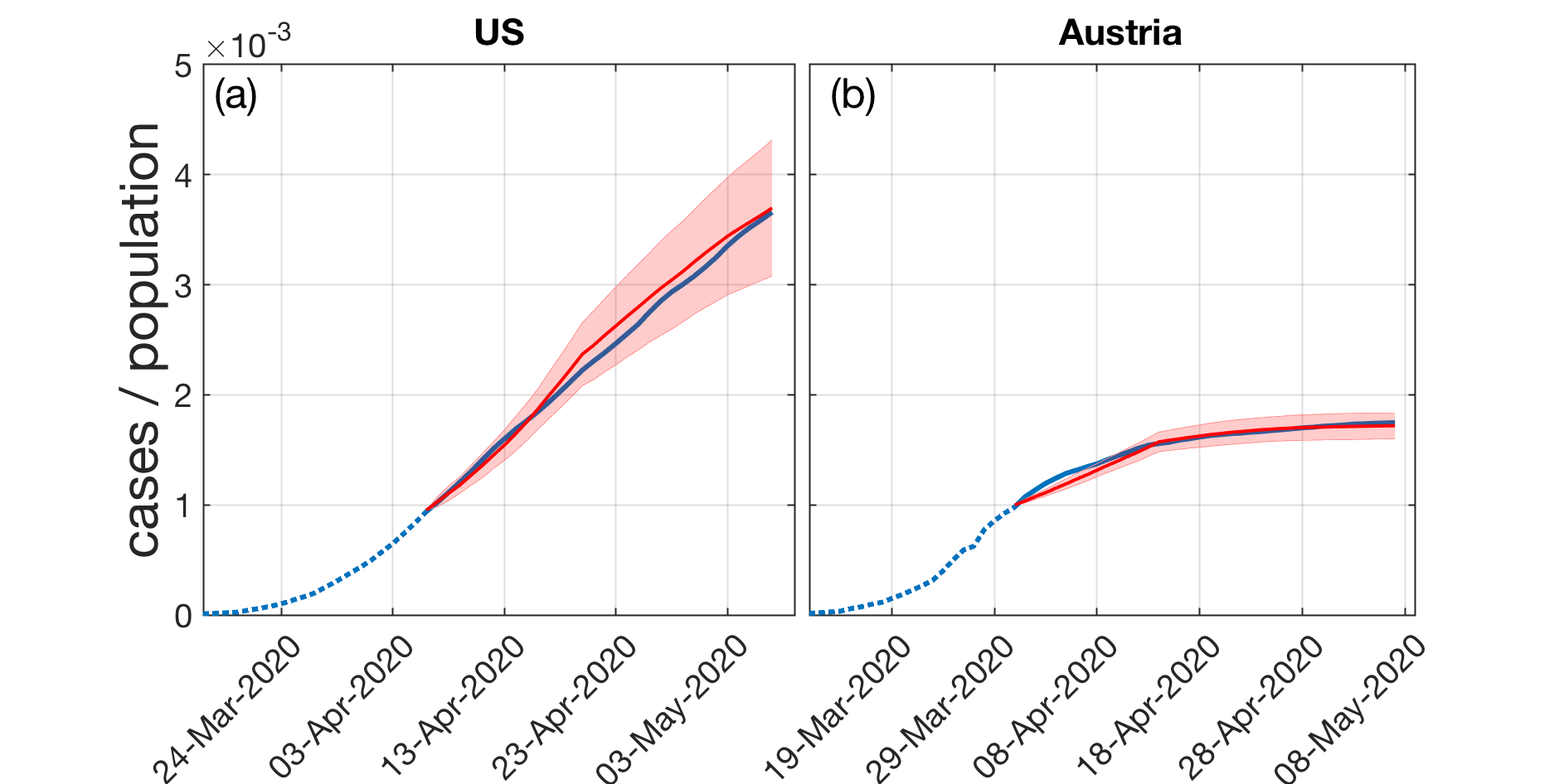} 
		\caption{Model infection curves (red)
		when calibrated to the COVID-19 curves of positively tested in 
		(a) the US and (b) Austria.	 Five realizations with different sets of initially infected are shown.
		The simulation starts when more than a tenth of a percent of the population was tested positive.
		The situation in the US assumes a Poissonian small-world network with average (daily) degree $D=5$. 
		The lockdown scenario in Austria that has been in place from March 16 to May 15, 2020 is modeled with 
		social contacts limited to households, $D=2.5$. For the choice of the other model parameters, see text. 
		The model clearly produces the correct type of infection curves.
	}
	\label{fig:calibration}
\end{figure}

{\bf Calibration.} 
Finally, we calibrate the model to the COVID-19 infection curves of two countries, the US and Austria. 
For this we have to take the following assumptions on the model parameters. 

The viral dynamics of COVID-19 is highly heterogenous \cite{heterogenious}. 
Motivated by evidence that people cary viral loads and are thus contagious for more than 
20 days after disease onset (most of them, however, being lower -- depending on severity) \cite{he,after_symp}, 
and given that infectiousness can start 2-3 days before showing symptoms \cite{he},  
we use $d=14$ days.  

In 2019 the average household size in the EU was 2.3 people \cite{eurostat}.  
If we assume that at work and during leisure activity on average one meets 3-4 people more per day, 
we decide to use an average degree of $D=5$ in our Poissonian small-world for normal conditions.
If we assume that on average about 30\% of all of our social relations are outside of our household, we 
set $\epsilon=0.3$. This is a somewhat arbitrary choice, however, 
note that deep in the linear regime, $\epsilon$ is found to be an almost irrelevant parameter that does not influence  
the outcome in significant ways. 
To model the lockdown that was imposed in Austria on Mar 16, 2020 we assume that its effect is 
basically to reduce social contacts to within households, and eliminate any other contacts. For this 
scenario we assume $D=2.5$, and $\epsilon=0$. 
Finally, for the daily transmission rate we set $r = 0.0149$. 
This choice is motivated by estimates of the COVID-19 individual-level secondary attack rate (SAR) 
in the household setting, which is reported at about 19\%, \cite{SAR}, and the relation $r=1-(1-SAR)^{1/d}$. 
Note that these estimates of the parameters are based on recent estimates (not yet peer reviewed) and might change 
in the future. Also note the SIR limit for this case  being at $R(t\to\infty)\sim$ 20\%, which is somewhat 
lower than what is expected in \cite{herd4,herd5}.
With these parameters we find a critical degree of $D_c=7.2$.
 
We use 100,000 nodes and 40 and 100 initially infected for the US and Austria, respectively.  
Since it is not possible to compute every individual in the simulation, 
we decided to initiate the simulation at the point where $0.1$\% of the population is tested positive that is   
Apr 7 for the US and Apr 3 in Austria. 
For the respective population sizes we use United Nations data from 2019 \cite{population2019}. 

In Fig. \ref{fig:calibration} we show the model infection curves in comparison to the number of 
positively tested persons \cite{JHU} for the US (a) and Austria (b). 
Solid blue lines mark the situation where more than $0.1$\% of the population was tested positive; 
simulations are performed  from that date on. 
Note that one case in the model represents many in reality. In the simulations 
relatively few cases are produced and the integer steps are still visible. 
This produces the wiggly appearance of the curves.
Obviously, the model produces infection curves of the observed type. 

\section{Discussion and Conclusions}
Here we offer an understanding of the origin of the extended linear region of the 
infection curves that is observed in most countries in the current COVID-19 crisis. 
This growth pattern is unexpected from mainstream epidemiological understanding. 
It can be understood as a consequence of the structure of low-degree contact networks 
and appears naturally as an hitherto unobserved (phase) transition from a 
linear growth regime to the expected S-shaped curves. 

We showed that for any given transmission rate there exists a critical degree of contact networks 
below which linear infection curves must occur, and above which the classical S-shaped curves
appear that are known from epidemiological models.  
The model proposed here is based on a simple toy contact network that mimics features of 
a heterogenous  degree, the small-work property, the fact that people tend to live in small groups that overlap and 
the fact that distant groups are linked through work and leisure activities.
We showed how the model can be used to simulate the effects of non-pharmaceutical interventions 
in response to the crisis by simply switching to low-degree networks that do not allow for 
linking of distant groups. 

The model not only allows us to understand the emergence of the linear growth regime, 
but also explains why the epidemic halts  much below the levels of herd immunity. 
Further, it allows us to explain the fact that in countries which are beyond the maximum of the 
epidemic, a relatively small number of daily cases persists for a long time. 
This is because small alterations and re-arrangements in the contact networks 
will allow for a very limited spread of infections. 

We find that for the empirically motivated parameters used here, the critical degree is $D_c=7.2$, which 
is above the degree of the contact networks for which we effectively assume $D\sim5$. 
This means that linear growth must be expected. 
Note that countries with larger family structures might be closer to the critical degree, above which 
catastrophic epidemic spreading would occur. 

The linear growth phase appears to be dominated by {\em cluster transmission} of the disease, 
meaning that new infections primarily appear in the ``small worlds'' or local network neighborhood 
(households, workplaces, etc.) of infected individuals. In the superlinear (exponential) phase, 
sustained {\em community transmission} sets in where new cases cannot be traced to 
already known cases in their neighborhood. In this regime, transmission across the shortcuts 
in the network becomes more prevalent. This effective mixing of the population gives a dynamic 
that approaches the mean field case of SIR-like models.

 Finally, we calibrated the model to realistic network parameters, transmission rates, and the time of being contagious 
and showed that realistic infection curves (examples of the US and Austria are shown) 
emerge without any fine-tuning of parameters. The onset of the 
lockdown -- and the associated reduction of the degree in the contact networks -- 
determines the final size of the outbreak which is well below the levels of herd immunity.   

Given the number of countries that enter linear growth phases,
our results raise serious concerns regarding the applicability of standard compartmental models 
to describe the containment phase that can be achieved by means of non-pharmaceutical interventions.
SIR-like models show linear growth only after fine-tuning parameters and linear growth would be a mere statistical fluke.
We argue that network effects must be taken into account to understand post-intervention epidemic dynamics.

\section{Methods}

{\bf Poissonian small-world network.}
For the network $A$ we use a Poissonian small-world network, which generalizes the usual regular 
small-world network in the sense that the degree is not fixed, 
but is chosen from a Poissonian distribution, characterized by $\lambda$.
The network is created by first imposing a Poissonian degree sequence on all nodes. 
Assume that nodes are arranged on a circle. Nodes are then linked to their closest neighboring nodes on the circle.   
This creates a situation where every person is member of a small local community. 
As for real families, these communities 
strongly overlap. 
Finally, as for the conventional small-world network, with probability $\epsilon$
we re-link the links of every node $i$ to a new, randomly chosen target node $j$, 
which can be far away in terms of distance on the circle. 
$\epsilon$ is the fraction of an individual's social contacts that are outside the local community (family). 
These links can be seen as links to colleagues at work or leisure activities, and allow us 
to model the existence of superspreaders \cite{Kupferschmidt}. 
Note that the actual average degree of the so-generated 
network is very close to the $\lambda$ of the Poisson distribution, $D\sim\lambda$.
We also implemented a conventional conventional small-world network with a fixed degree. 
When results are compared with the Poissonian small-world network
only marginal differences are observed. 

{\bf Order parameter.}
To distinguish the linear from the  sigmoidal growth, we propose a simple 
``order parameter'' as the standard deviation of all new daily cases 
(after excluding all days where there are no new cases), 
\begin{equation}
{\cal O} = {\rm std} (C(t)) \, . 
\end{equation}
Clearly,  for a linear growth of the infection curve, daily cases, $C(t)$, are constant, and the standard deviation is zero. 
For the S-shape growth, daily cases first increase then decrease over time, 
and the standard deviation becomes larger than zero. 
Hence, a standard deviation deviating from zero signals the presence of a non-linear increase 
of the cumulative positive cases, $P(t)$. 

\begin{acknowledgments}
\subsection{ACKNOWLEDGEMENTS}
We thank Christian Diem for helpful discussions. 
This work was supported in part by the Austrian Science Promotion Agency, 
FFG project under  857136, 
the Austrian Science Fund FWF under P29252, 
the WWTF under COV 20-017, and the 
Medizinisch-Wissenschaftlichen Fonds des B\''urgermeisters der Bundeshauptstadt Wien under CoVid004.   
\end{acknowledgments}

 


\newpage 
.
\pagebreak 
 
\onecolumngrid
\appendix

\section{Supplementary Information}

\subsection{the SIR model}

The classic epidemiological model is the SIR model that models the time evolution of 
compartments of susceptible S(t), infected I(t), and recovered R(t) individuals. 
It can be formulated as the set of non-linear differential equations, 
\begin{eqnarray}
\frac{d}{dt} S &=& -\beta \frac{SI}{N} \nonumber \\
\frac{d}{dt} I &=& \beta \frac{SI}{N} - \gamma I  \quad . \\
\frac{d}{dt} R &=& \gamma I  \nonumber 
\end{eqnarray}
At any point in time, $S+I+R=N$. To solve the equations, for a given $N$ 
the initial conditions, $S(0)$ and $I(0)$ must be specified. 
$\gamma$ is called the recovery rate, 
$\beta$ is the infection rate, controlling how often a susceptible--infected contact results in a new infection.  
To play with it online, see e.g. \url{http://www.public.asu.edu/~hnesse/classes/sir.html?Alpha=0.3&Beta=0.1666&initialS=990&initialI=10&initialR=0&iters=70}

\subsection{Countries in early phase of the pandemic}

\begin{figure}[htb]
	\centering
	 \includegraphics[width=0.50\columnwidth]{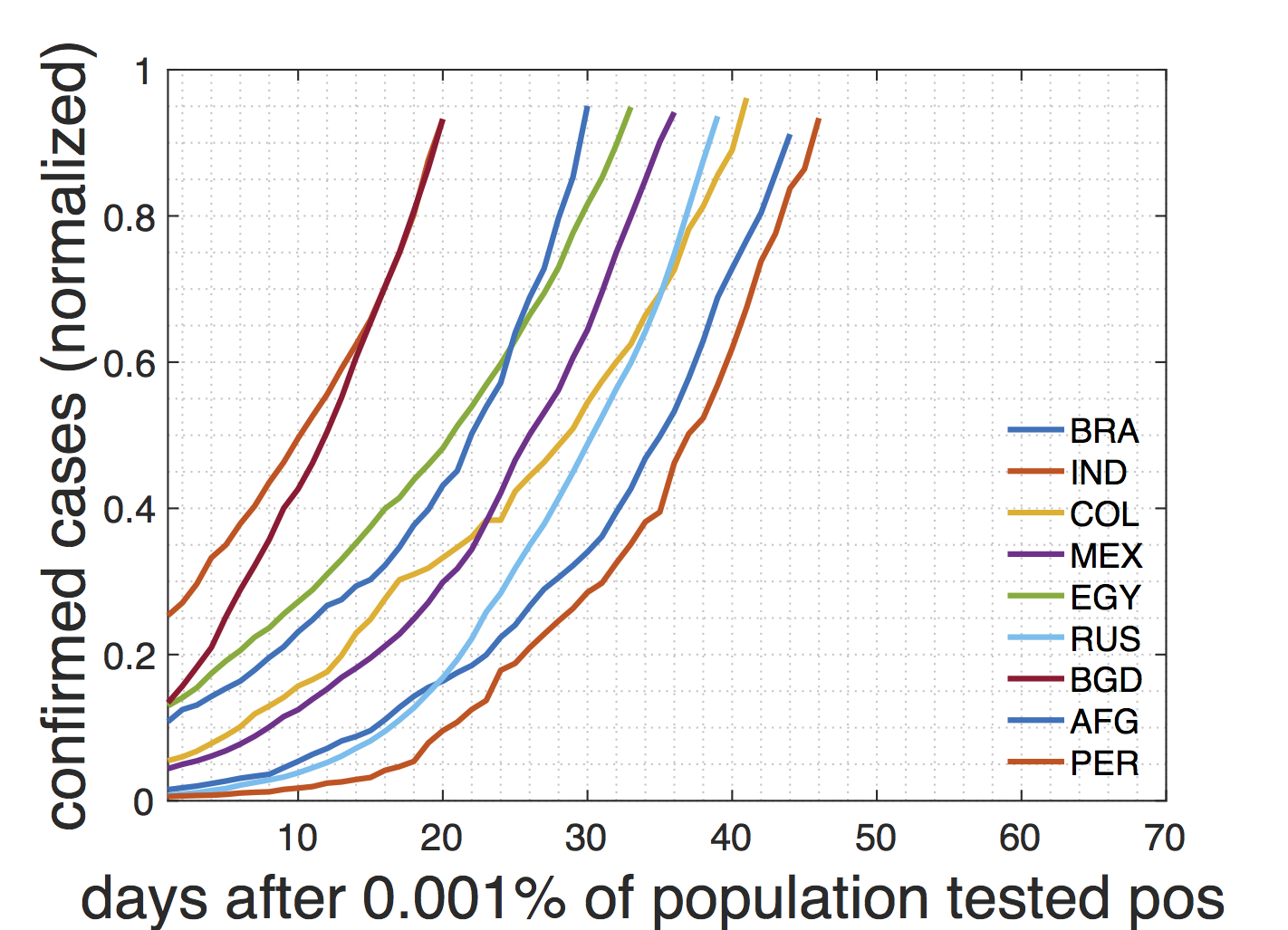}
	\caption{Cumulative numbers of positively tested cases, as in Fig. \cite{fig:linear},  normalized to the last day (May 8, 2020). 
	for countries with an exponential growth, possibly because they are still in an early phase.  
	}
	\label{SIfig:linear}
\end{figure}

In SI Fig. \ref{SIfig:linear} we show several countries that display the typical initial exponential growth 
of the infection curve. Many of these countries are in South America.  

\subsection{Position of critical degrees for various parameter settings}

\begin{figure}[h]
	\centering
	 \includegraphics[width=0.7\columnwidth]{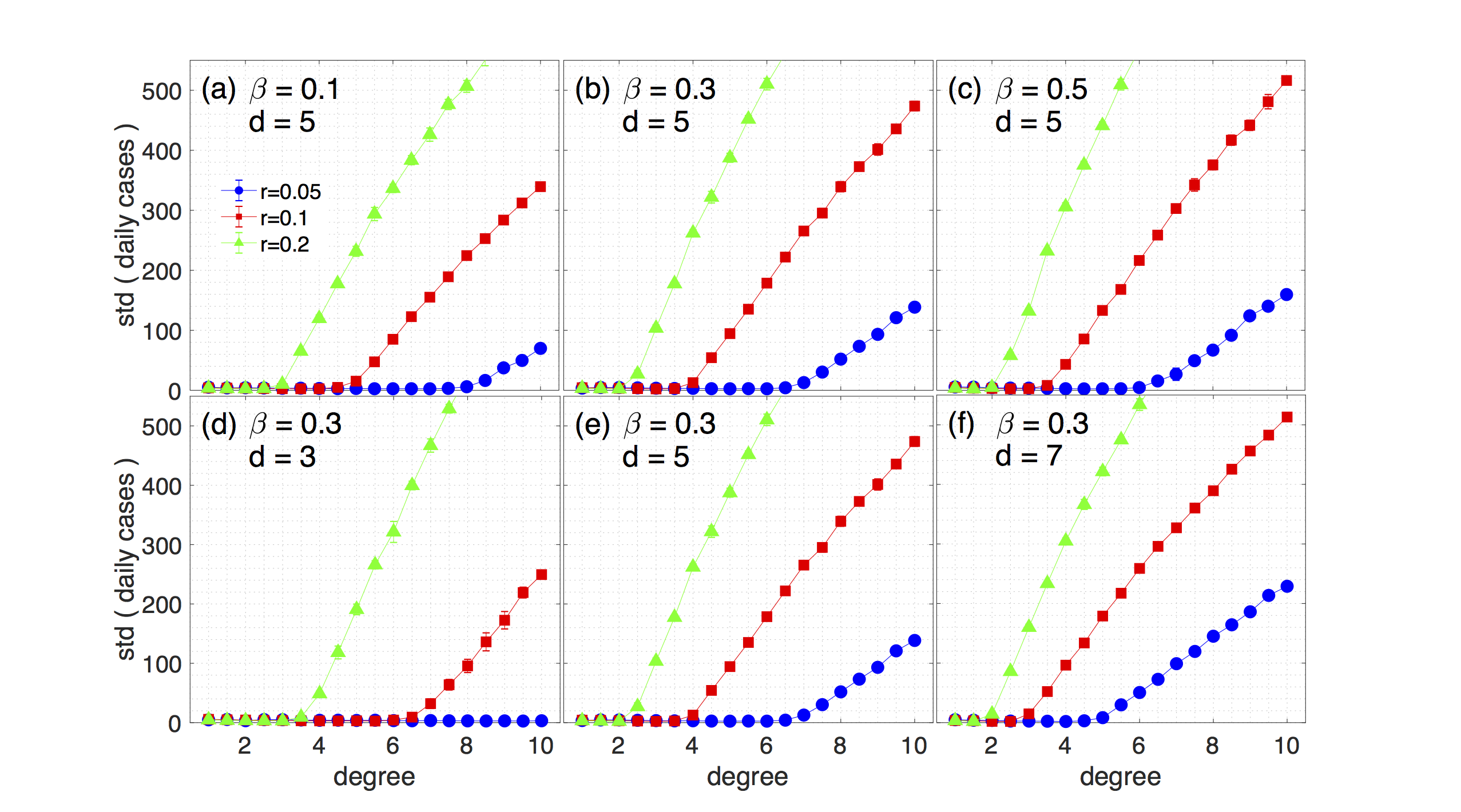} 
	\caption{Transitions from linear growth (${\cal O}\sim0$) 
	based on a Poissonian  small-world networks with $N=10,000$ with degrees ranging from 0 to 10.
	Cases for transmission probabilities of $r=0.05$,  $0.1$, $0.2$, and for rewiring probabilities  
	$\epsilon=0.1$, $0.3$, and $0.5$ are shown. Days of being contagious are $d=2$, $d=4$, and $d=6$. 
	}
	\label{SIfig:X}
\end{figure}

In SI Fig. \ref{SIfig:X} we see the position of the transition from linear to traditional epidemic growth 
in terms of critical degrees. The order parameter {\cal O} is plotted on the y-axis. In  traditional
sigmoidal growth {\cal O}  takes values well above zero, the onset defines the critical degree,$D_c$, for the 
various parameter settings, collected in Table 1. 
The position of critical degrees is practically the same with a network of size $N=1000$, 
assumed that both are initialized with 10 initially infected.

\subsection{Critical degrees for various parameter settings on a small-world network}

\begin{figure}[h]
	\centering
	 \includegraphics[width=0.24\columnwidth]{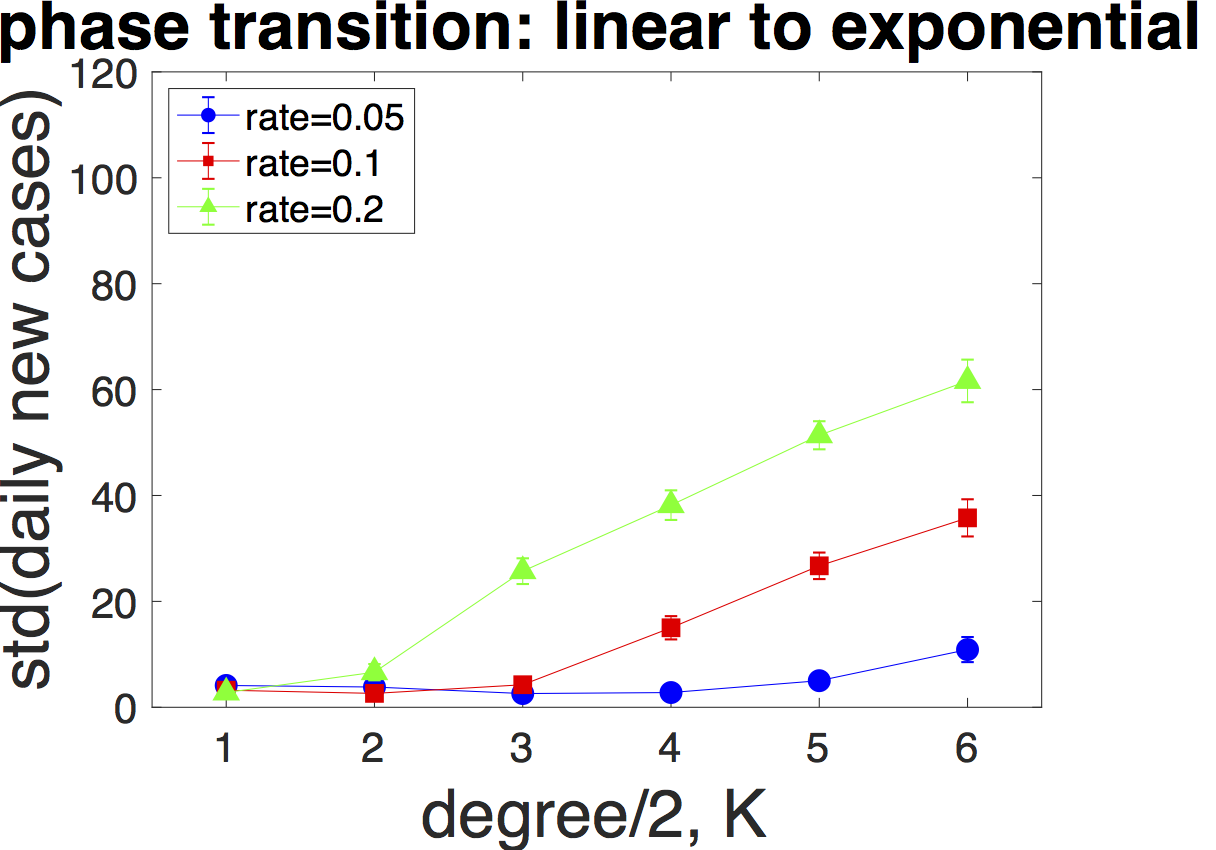} 
	 \includegraphics[width=0.24\columnwidth]{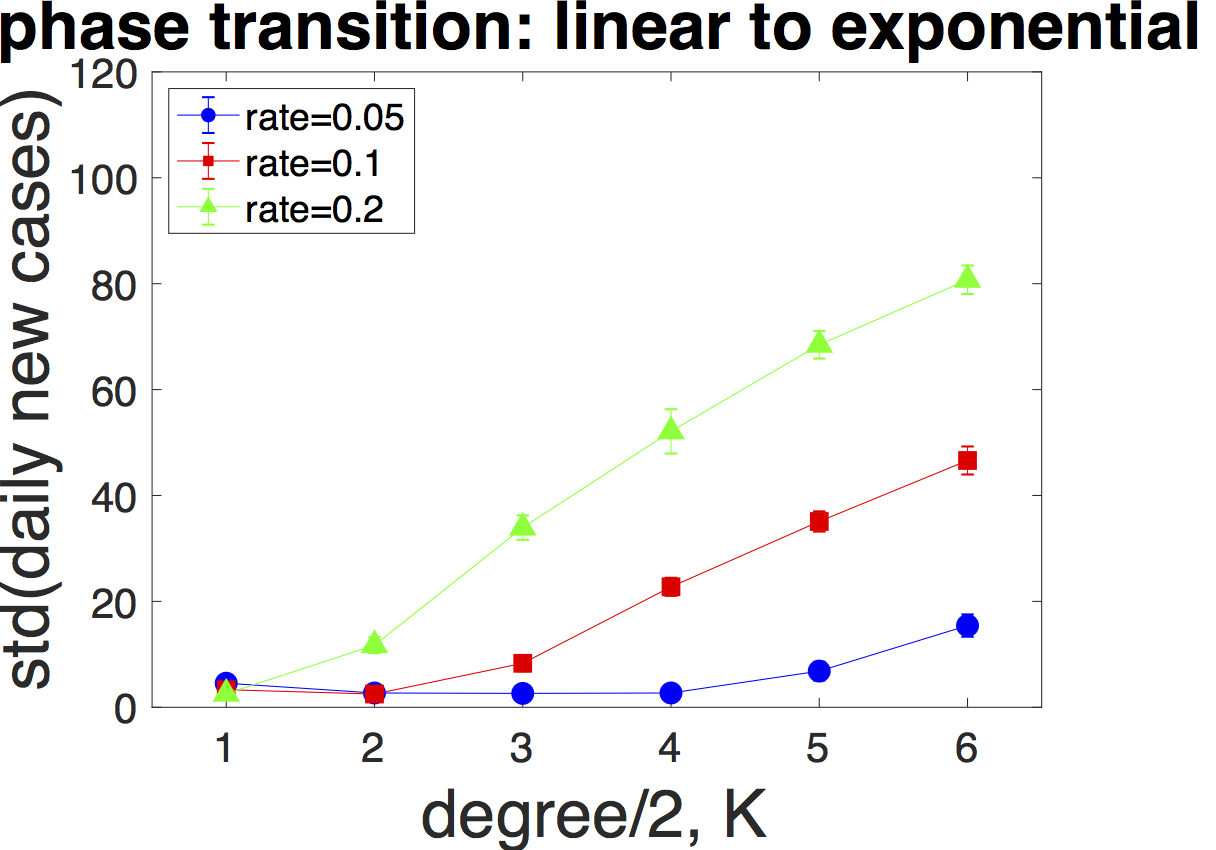} 
	 \includegraphics[width=0.24\columnwidth]{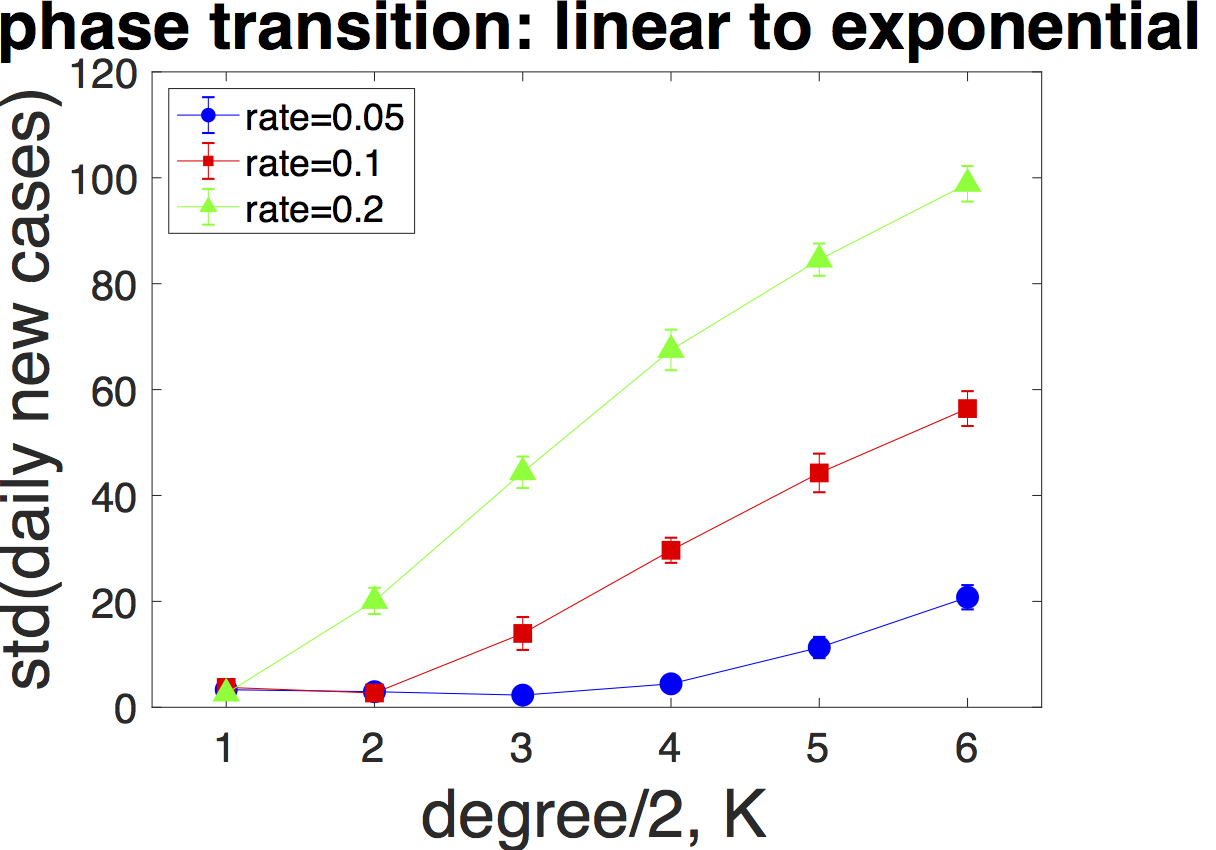} 
	 \includegraphics[width=0.23\columnwidth]{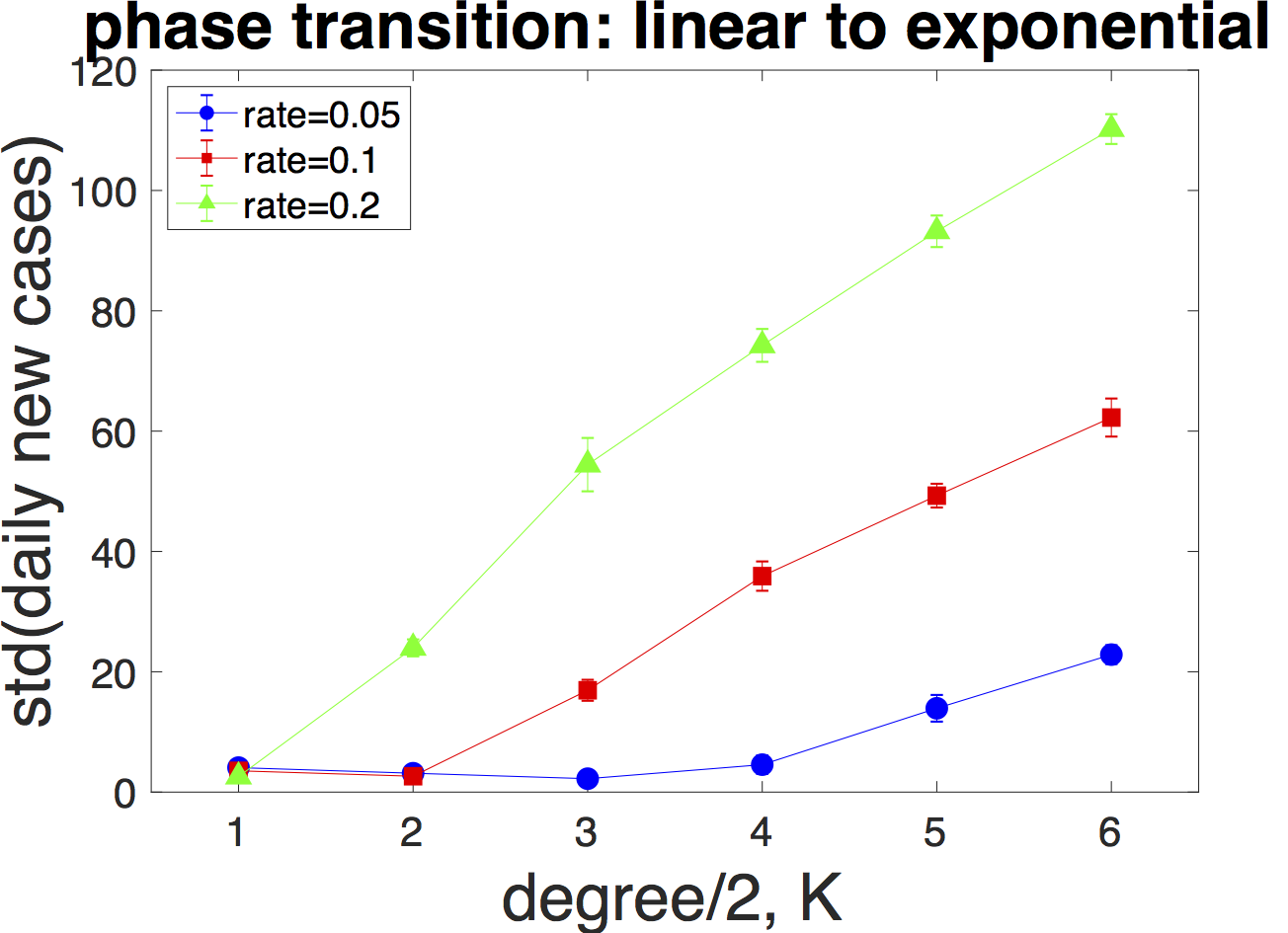} \\
	 \includegraphics[width=0.24\columnwidth]{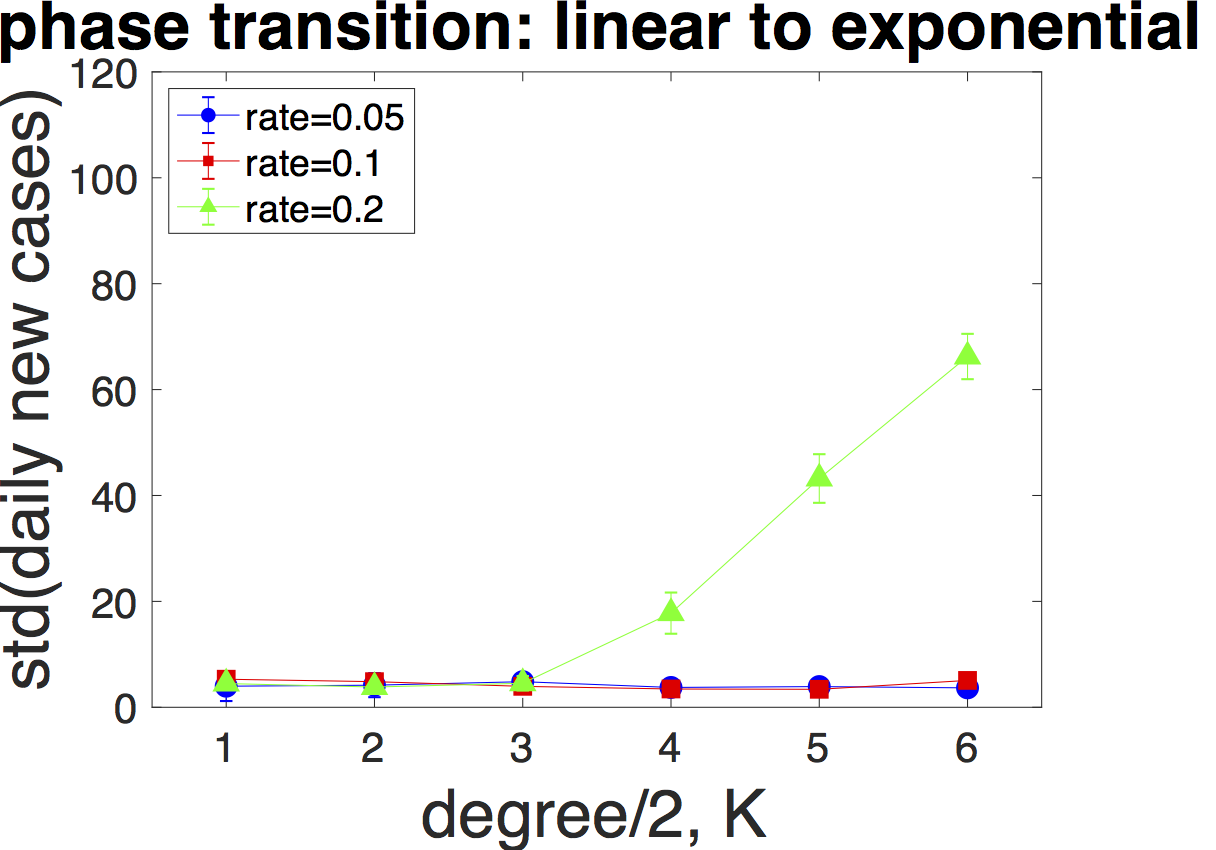} 
	 \includegraphics[width=0.24\columnwidth]{figures/SIphase_trans_beta03.png} 
	 \includegraphics[width=0.24\columnwidth]{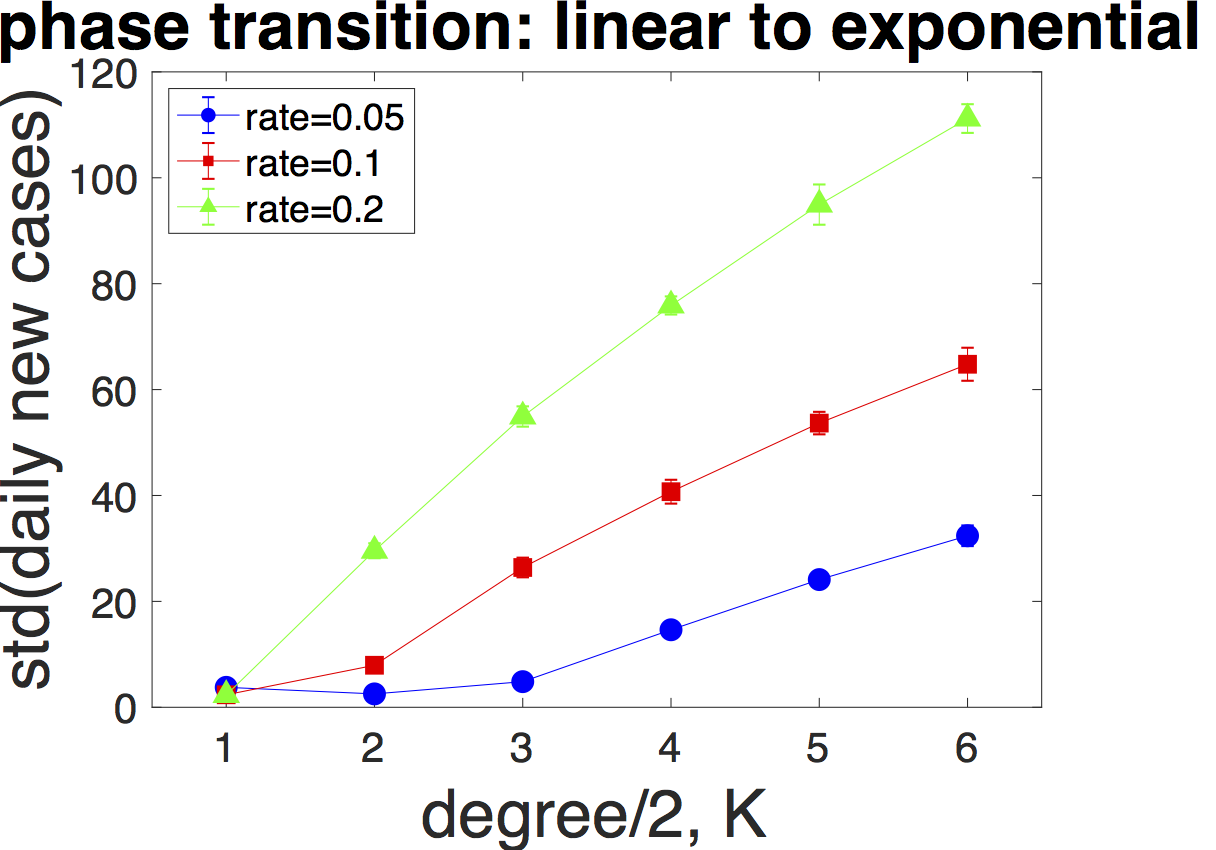} 
	 \includegraphics[width=0.24\columnwidth]{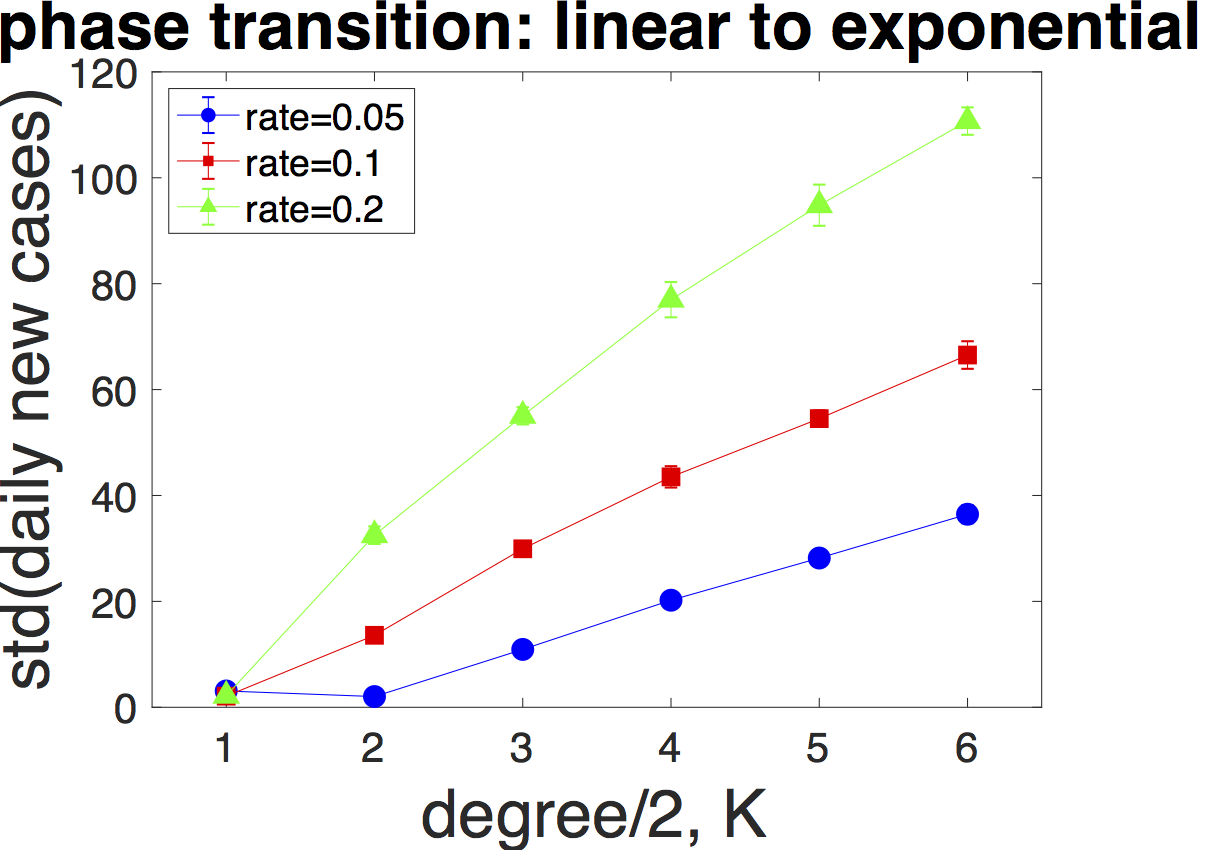} 
	\caption{Position of  critical degrees for a standard small-world networks with $N=1000$,
	with degree $D=2K$ and transmission rates of $r=0.05$,  $0.1$, $0.2$.
	(top) Rewiring parameters shown are $\epsilon=0.05$, $0.1$, $0.2$, $0.3$. 
	for $d=5$ days being infectious. 
	(bottom) Here we set	$\epsilon=0.3$ and vary the days being contagious: $d=1$, $d=4$, $d=6$, $d=9$.
	In all cases 10 randomly chosen nodes were initially infected. 
	}
	\label{SIfig:SW}
\end{figure}

The positions of the critical degrees (order parameter becoming larger than 0) 
are shown for a standard small-world networks 
with $N=1000$ in SI Fig. \ref{SIfig:SW}. Note that here we have used an even degree, 
indicated by $2K$. 

\end{document}